\documentclass[useAMS,usegraphicx,times]{mn2e}
\usepackage{mathptmx}
\def\H0{{\it H}$_0$}
\def\Ms{{\it M}$_\odot$}
\def\Ls{{\it L}$_\odot$}
\def\kmps{km~s$^{-1}$}
\def\ergps{erg~s$^{-1}$}
\def\kmpspMpc{km~s$^{-1}$~Mpc$^{-1}$}
\def\Ms{{\it M}$_\odot$}

\def\nH{$N_{\rm H}$\thinspace} 

\def\psqcm{cm$^{-2}$}
\def\ergpspsqcm{erg~cm$^{-2}$~s$^{-1}$}

\def\cps{ct\thinspace s$^{-1}$}

\def\phpspsqcm{ph\thinspace s$^{-1}$\thinspace cm$^{-2}$}

\title[XMM-Newton observations of IRAS F15307+3252] 
{X-rays from the Type II quasar in the hyperluminous infrared galaxy IRAS F15307+3252}
\author[K. Iwasawa et al]
{\parbox[]{6.5in} {K. Iwasawa$^1$, C.S. Crawford$^1$, A.C. Fabian$^1$ and R.J. Wilman$^2$}\\
  \\
  $^1$Institute of Astronomy, Madingley Road, Cambridge CB3 0HA\\
  $^2$Department of Physics, University of Durham, South Road, Durham DH1 3LE\\
} 
\date{}

\begin{document}

\maketitle
\begin{abstract}
  We report the detection of X-ray emission from the hyperluminous
  infrared galaxy IRAS F15307+3252 at $z=0.93$ and its properties
  obtained from XMM-Newton observations. Although the X-ray emission
  is very faint and the data are noisy, a prominent line-like feature
  in the observed 3--4 keV range is inferred from both photometric and
  spectroscopic techniques. It indicates an X-ray spectrum dominated
  by 6.4 keV Fe K$\alpha$ emission and the presence of a Compton-thick
  AGN. Our estimate of the luminosity of the illuminating source
  ($L_{\rm 2-10keV}\geq 1\times 10^{45}$\ergps), required to produce
  the observed Fe K$\alpha$ emission in reflection from cold matter,
  means that the hidden quasar nucleus accounts for a significant
  fraction of the large bolometric luminosity. The soft X-ray emission
  below 2 keV is found to be spatially extended and probably of a
  separate origin. The temperature and bolometric luminosity
  ($kT\simeq 2$ keV and $L_{\rm bol}^{\rm CL}\simeq 1\times
  10^{44}$\ergps) obtained from a thermal spectrum place this X-ray
  source on the $L$-$T_{\rm X}$ relation of galaxy clusters. The
  possible association with a galaxy cluster can be added to the list
  of remarkable similarities between IRAS F15307+3252 and another
  hyperluminous infrared galaxy IRAS 09104+4109 ($z=0.44$), both of
  which have bolometric luminosities dominated by hidden quasar nuclei.
  Our result on IRAS F15307+3252 illustrates how difficult it is to
  detect Compton thick Type II quasars at $z =1$, particularly if
  their bolometric outputs do not rival the hyperluminous population.
\end{abstract}


\section{Introduction}

The class of hyperluminous infrared galaxies, whose bolometric
luminosities exceed $10^{13}$\Ls, have been found through far-infrared
and submillimetre surveys by IRAS and SCUBA (e.g., Rowan-Robinson
2000), and more are expected from the Spitzer Space Telescope. Some of
them may be massive galaxies in formation at high redshift and, given
the suggested link between galaxy spheroids and central black holes in
them (e.g., Tremaine et al 2002), they might represent an important
galaxy evolution phase where a quasar nucleus is forming amid heavy dust
obscuration (e.g., Sanders et al 1988). A population of
mid-infrared-selected hyperluminous objects at $z>2$, which appear to
be powered mainly by obscured active galactic nuclei (AGN), are
emerging from Spitzer surveys (Houck et al 2005). As well as the
origin of the enormous luminosity being of a great interest,
examining the role of luminous counterparts of nearby Compton-thick
Seyfert AGN to the X-ray background (XRB) and
their detectability is also important, since the higher energy part of
the XRB is not yet fully resolved (Worsley et al 2005).

IRAS F15307+3252 is one of the few classical IRAS-selected hyperluminous
infrared galaxies at redshift of $z=0.926$ (Cutri et al 1994). With
the currently popular cosmology with $H_0 = 70$ \kmpspMpc,
$\Omega_{\rm M}=0.27$, and $\Omega_{\Lambda}=0.73$, the luminosity
distance is 6.1 Gpc and and the bolometric luminosity is estimated to
be $3\times 10^{13}$\Ls.

The large bolometric luminosity was once suspected to be the result of
gravitational lensing (Liu, Graham \& Wright 1996) as in the case of IRAS
F10214+4724 at $z=2.3$ (e.g., Broadhurst \& Lehar 1995). However
Farrah et al (2002) have ruled this out based on HST
imaging. Therefore, IRAS F15307+3252 indeed belongs among the
most luminous objects in the universe.

The UV/optical emission-line spectrum of IRAS F15307+3252 is of
Seyfert 2 type (Cutri et al 1994; Soifer et al 1994; Liu et al 1996;
Evans et al 1998). Optical spectropolarimetry has revealed broad MgII
$\lambda 2798$ emission and a rest-frame UV spectrum which is very
similar in shape and intensity to that of typical quasars, indicating the
presence of a dust-shrouded quasar nucleus (Hines et al 1995).

Despite the optical evidence for hidden active nuclei, hyperluminous
infrared galaxies are generally very faint in X-ray (Wilman et al
1998; 2003; see Franceschini et al 2000; Iwasawa, Ettori \& Fabian
2000 for IRAS P09104+4101; Iwasawa 2001; Alexander et al 2005 for IRAS
F10214+4724). IRAS F15307+3252 is no exception and previous X-ray
observations with ROSAT and ASCA failed to detect any X-rays (Fabian
et al 1996; Ogasaka et al 1997). We report here the detection of faint
X-ray emission from this hyperluminous infrared galaxy and its
properties obtained from XMM-Newton observations, and discuss their
implications.

\section{Observations}

IRAS F15307+3252 was observed with the EPIC cameras of XMM-Newton on
three occasions (Table 1). Both EPIC pn and MOS cameras operated in
the full-window mode with the medium filter in all the three
observations. A significant fraction of the datasets are affected by
high radiation background. A larger fraction of the observed intervals
are affected in the pn data than in the MOS data. Time intervals of
quiescent background (count rate less than 0.3 \cps\ for each of the
MOS and less than 1.2--1.5 \cps\ for the pn) were selected, based on
the single event light curves at energies above 10 keV taken from the
whole field of each detector. The useful exposures from the three
observations are 6, 10, and 5 ks (21 ks in total) for the pn camera,
and 10, 12, and 11 ks (33 ks in total) for the MOS cameras. Single and
double events recorded by the detectors were used for the data
analysis presented here. The data reduction was performed by the
standard XMM-Newton software package SAS 6.1 and LHEASOFT 5.3.1.


\begin{table}
\begin{center}
  \caption{XMM-Newton observations of IRAS F15307+3252. Useful
    exposure times in unit of ks (and whole duration of exposure
    including the high background intervals in parenthesis) for the
    respective EPIC cameras.}
\begin{tabular}{lcc}
Date & pn & MOS \\
& ks & ks \\[5pt]
2002 Jul 30 & 6.7 (24) & 10 (25)\\
2004 Jul 29 & 9.6 (30) & 12 (31) \\
2004 Dec 30 & 5.1 (23) & 11 (18)\\
\end{tabular}
\end{center}
\end{table}

\section{Results}

\subsection{Photometric study}

IRAS F15307+3252 is detected with XMM-Newton but is very faint,
particularly at energies above 2 keV. With so few detected counts, the
individual observations do not warrant an independent analysis. We
integrated all three datasets to examine source detection first in
different energy bands. Compton-thick obscuration can be a plausible
explanation for the X-ray faintness of IRAS F15307+3252. Since a
strong Fe K$\alpha $ feature at rest-frame 6--7 keV would characterise
the X-ray spectrum in such a case, we designed a photometric study by
dividing the whole bandpass into the following four bands: 0.5--2 keV,
2--3 keV, 3--4 keV, and 5--10 keV, so that Fe K$\alpha $ feature falls
in the 3--4 keV band by the galaxy redshift ($z=0.926$). X-ray images
in the four bands were constructed by adding all the pn and MOS data
from the three observations, which are shown in Fig. 1.

IRAS F15307+3252 is detected at more than $3\sigma $ in all but the
2--3 keV band images (see Table 1). The source is strongly detected at
lower energies below 2 keV for which further details are presented
below. This component does not seem to stretch well above 2 keV, as
suggested by the lack of detection in the 2--3 keV band. Note, however, a
remarkable recovery of the source in the 3--4 keV band. This indicates
a strong excess in the 3--4 keV range, where Fe K$\alpha $ emission is
expected. The 2--3 keV drop-out and the marginal detection in the
5--10 keV band suggest a rather hard underlying continuum in the 2--10
keV (the rest frame 4--20 keV) range. These are characteristics of the
X-ray spectrum of a heavily absorbed active nucleus.


\begin{figure}
\centerline{\includegraphics[width=0.4\textwidth,angle=0,
    keepaspectratio='true']{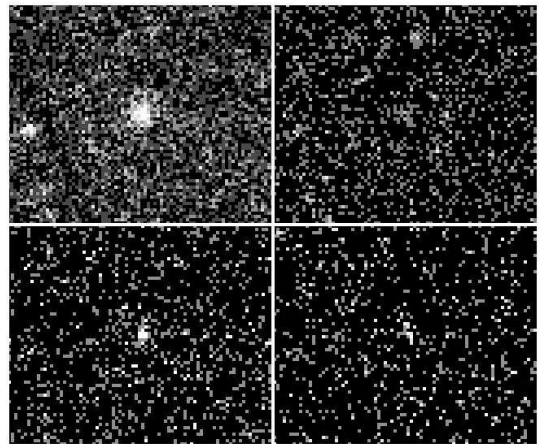}} 
  \caption{ The X-ray images in the four energy bands: clockwise from
    upper-left: 0.5--2 keV; 2--3 keV; 5--10 keV; and 3--4 keV. The
    images were constructed by adding all the pn and MOS detectors
    together from all the three observations. The size of each image
    is $5^{\prime}\times 4^{\prime}.5$. The background-corrected
    counts in the respective bands and their signal to noise ratios
    (in parenthesis) are {\bf 0.5--2 keV:} 186 ct ($12\sigma$); {\bf
      2--3 keV:} 11 ct ($2\sigma$); {\bf 3--4 keV:} 34 ct ($5\sigma$);
    and {\bf 5--10 keV:} 33 ct ($4\sigma$).}
\end{figure}


\subsection{Extended soft X-ray emission and its origin}

\begin{figure}
\centerline{\includegraphics[width=0.3\textwidth,angle=270,keepaspectratio='true']{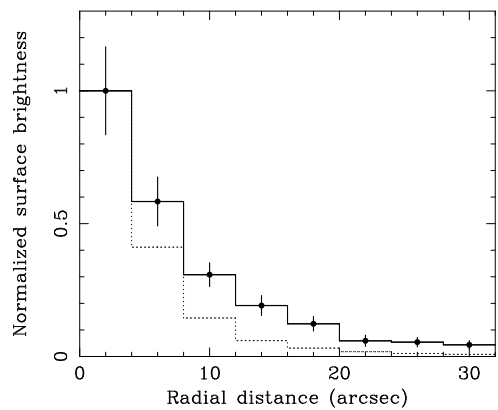}}
\caption{ The radial X-ray surface brightness profiles of the 0.5--2
  keV emission from IRAS F15307+3252 (solid-line histogram) as
  observed with XMM-Newton, together with that from NGC4395 (dotted-line histogram) in the
  same band. The profiles have been normalized at the innermost bin.  
NGC4395 is a point-like source thus represents the point
  spread function.}
\end{figure}

\begin{figure*}
\hbox{\hspace{12mm}{\includegraphics[width=0.41\textwidth,angle=0,keepaspectratio='true']{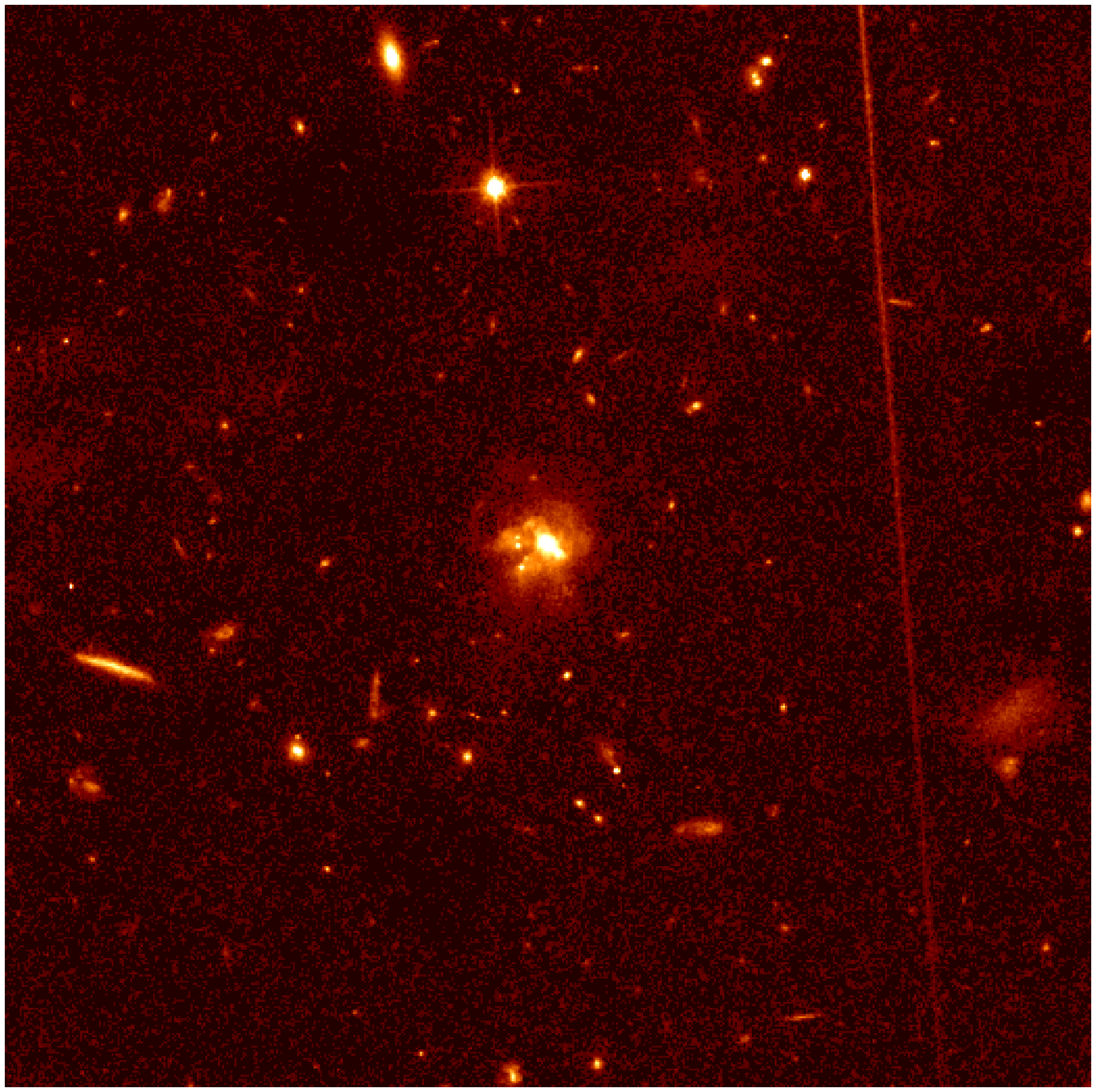}}
\hspace{8mm}
{\includegraphics[width=0.406\textwidth,angle=0,keepaspectratio='true']{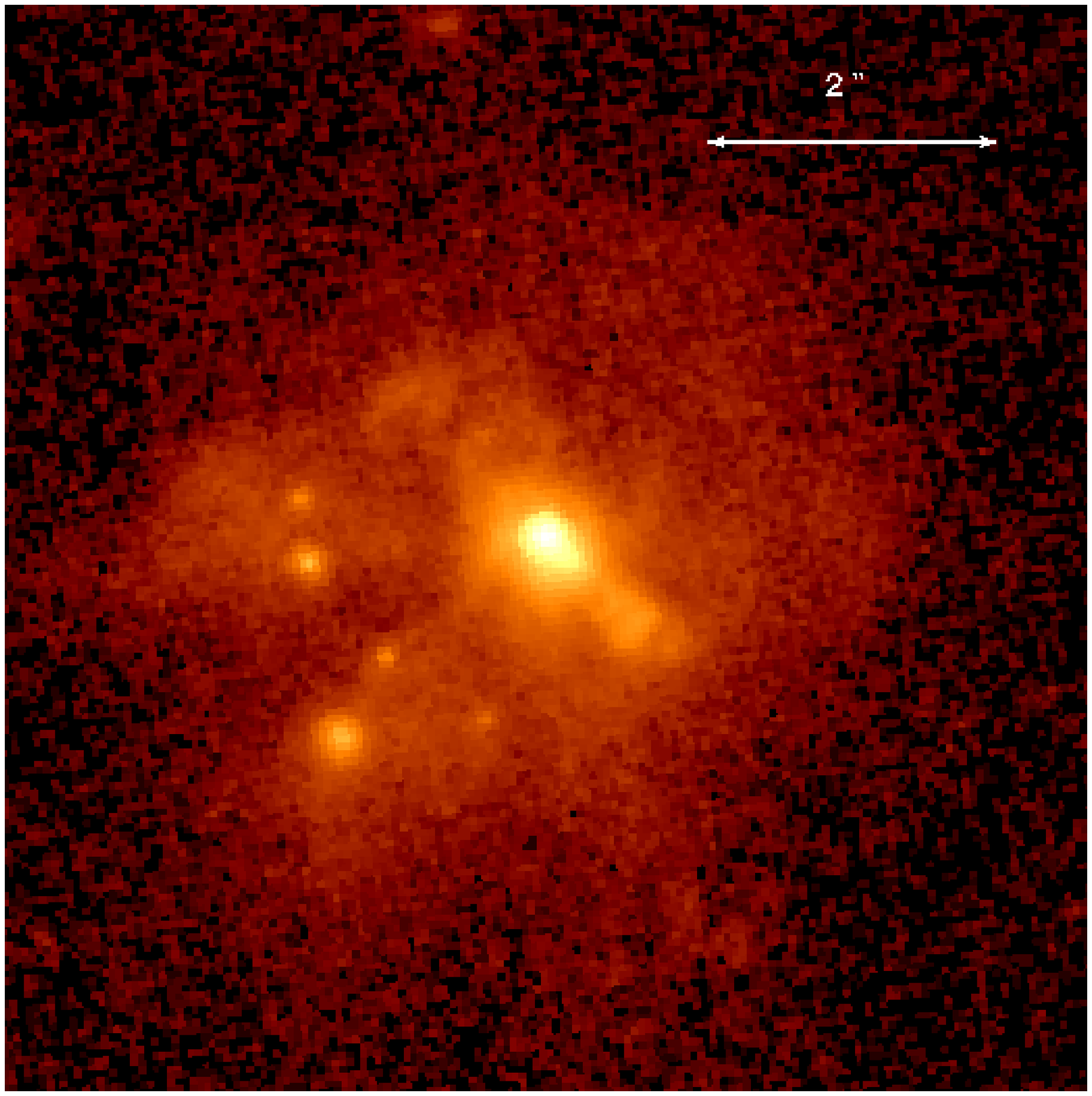}}}
\caption{The HST/ACS images of the $1^{\prime}\times 1^{\prime}$
  region (left) around IRAS F15307+3252 (right), taken through the
  F814W filter (rest-frame blue). North is up and East to the left.
  The images have been gaussian-smoothed by 1 pixel ($\simeq
  0^{\prime\prime}.05$). The large-scale image shows a chain of small
  galaxies 11$^{\prime\prime}$--15$^{\prime\prime}$ ($\sim 100$ kpc at
  $z=0.93$) to the south of the IRAS galaxy, which is reminiscent to
  Markarian's chain of galaxies in the Virgo cluster lying at $\sim
  300$ kpc in projected distance from M87. The zoom-up of IRAS
  F15307+3252 shows complex structures within the diffuse envelope,
  including several knots, a cavity of a conical shape with a sharp
  boundary to the south-east, and arc-like structures to the north-east.}
\end{figure*}

The soft X-ray emission below 2 keV found at the position of IRAS
F15307+3252 is not point-like. The radial profile of the soft band
source is compared with that of the nuclear source in NGC4395, which
was also observed with the XMM-Newton in the full-window mode (Fig.
2). Since the X-ray emission from NGC4395 has been found to be
point-like as viewed by the Chandra
ACIS-S (Moran et al 2005), it represents the point spread function
(PSF). Although adding different detectors and observations together
might have resulted in a blurred PSF at some degree, a comparison of
the radial profile with other point sources in the field confirms the
extended nature of the soft X-ray emission from IRAS F15307+3252.

The angular scale at $z=0.926$ with the adopted cosmology is $\simeq
8$ kpc arcsec$^{-1}$. Detection of a source size broader than the PSF
(FWHM $\simeq $5-6 arcsec) means that the source is extending at
least $\sim 50$ kpc. Fitting a Lorentzian function to the respective
radial profiles of IRAS F15307+3252 and NGC4395 implies that the FWHM
of the intrinsic source extension is about $21\pm 5$ arcsec,
corresponding to 85 kpc in radius. This is larger than the effective
radius ($\sim 12$ kpc, Farrah et al 2002) of the giant elliptical host
of IRAS F15307+3252, of which the surface brightness distribution is
described well with the de Vaucouleur's profile (see also Liu et al
1996).

No galaxy cluster has so far been reported for the region around IRAS
F15307+3252. However, IRAS F15307+3252 is a luminous ($M_I=-26.4$,
Farrah et al 2002) giant elliptical, inside which some substructures
have been found, indicating galaxy mergers or interaction (Fig. 3,
also Soifer et al 1994; Liu et al 1996; Farrah et al 2002). The galaxy
is $\sim 1.5$ mag more luminous than the Virgo brightest galaxy M87,
and should be in a relatively massive dark matter halo. Fig. 3 shows
the HST/ACS image of the region around IRAS F15307+3252, retrieved
from the HST archive at STScI. The data were taken on 2002 August 18
in a 2,120-s exposure, using the F814W filter (approximately $I$
band). While there is no clear evidence for a galaxy concentration
centred on the IRAS galaxy as such expected in a rich cluster, a chain
of several small galaxies 11--15 arcsec ($\sim 100$ kpc) to the south
is seen, which lies within the X-ray extension, indicating a moderate
galaxy over-density. The HST image also shows that both IRAS
F15307+3252 and the several nearby compact knots (which Farrah et al
2002 ascribe to the remnants of merging companions) are embedded in an
envelope of diffuse emission out to a radius of $\sim$3 arcsec. There
is considerable structure within this emission, including a sharp
conical-shaped cavity to the SE with its apex at the AGN. Further
arc-shaped structures concentric to the active nucleus and
perpendicular to the cavity are also apparent, which could be due to
[OII]$\lambda 3727$ located near the edge of the filter bandpass,
representing ionization cones.

Fitting jointly the pn and MOS spectra (Fig. 4) gives a temperature of
$kT = 2.1^{+0.6}_{-0.4}$ keV (hereafter, the errors for spectral
parameters are of the 90 per cent confidence region for one parameter
of interest).  The 0.4--2.8 keV band was fitted with half solar
metallicity and Galactic absorption of \nH $=2\times 10^{20}$\psqcm\
(Dickey \& Lockman 1990) being assumed. The observed 0.5--2 keV flux
is $1.2\times 10^{-14}$ \ergpspsqcm. The absorption-corrected
bolometric luminosity is estimated to be $\sim 1\times 10^{44}$\ergps.
The temperature and bolometric luminosity are on the $L$-$T_{\rm X}$
relation for galaxy clusters, groups and elliptical galaxies (e.g.,
Fukazawa et al 2004), and IRAS F15307+3252 lies in the region for a
poor cluster (it is comparable to the Virgo cluster in temperature and
luminosity). We tentatively identify the extended soft X-ray emission
with hot gas associated with a relatively poor cluster around the IRAS
galaxy.

Whilst soft X-ray emission in some hyperluminous infrared galaxies has
been ascribed to star formation (Wilman et al 2003; Alexander et al
2005), the large source size, high temperature and large soft X-ray to
infrared luminosity ratio suggest that a starburst is unlikely to be a
major component of the soft X-ray emission in IRAS F15307+3252 (see
also Section 4.2). The cluster emission has a relatively short
radiative cooling time of $\leq 4$ Gyr. This would lead to a cooling
flow of $\sim 150 M_{\odot}$ yr$^{-1}$ if there is no heating to
balance cooling (there is a 6 mJy FIRST radio source at 1.4 GHz;
Becker, White \& Helfand 1995).


\subsection{Fe K line and hard X-ray emission}

\begin{figure}
\centerline{\includegraphics[width=0.34\textwidth,angle=270,keepaspectratio='true']{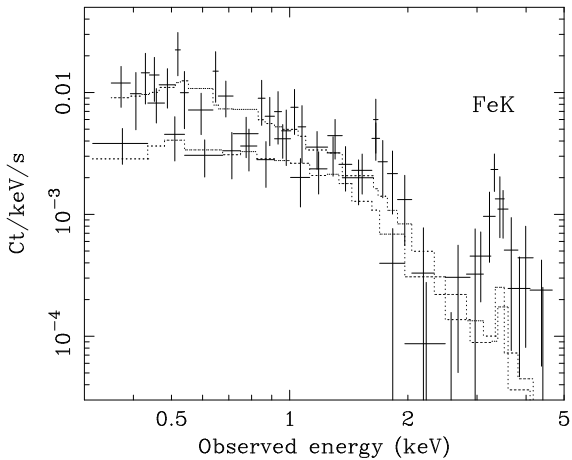}}
\caption{ The co-added pn and MOS spectra of IRAS F15307+3252. The
  energy scale is as observed. A strong excess feature in the 3--4 keV
  range can be identified with Fe K$\alpha$. The dotted-line
  histograms are the thermal emission spectrum (MEKAL) with $kT=2.1$
  keV, half solar metallicity and the Galactic absorption,
  best-fitting the 0.4--3 keV data. Note that the extrapolation of the
  thermal emission spectrum to higher energy clearly underestimates
  the observed Fe K emission flux, which is therefore of the different
  origin.}
\end{figure}

The photometric study (Section 3.1) demonstrated that X-ray emission
of IRAS F15307+3252 shows a strong excess in the 3--4 keV range, which
is probably due to a redshifted Fe K$\alpha $ line. There is a very
faint underlying continuum, which extends to higher energies. The
faint 5--10 keV band is too noisy for a spectral study and only fit
for a photometric measurement. The spectral data in the 2--5 keV band
can be fitted with a narrow gaussian with a flat power-law
($\Gamma<0.8$).  The centroid of the gaussian is
$6.47^{+0.10}_{-0.11}$ keV in the rest frame. The fit shows a line
width of 0.1 keV in gaussian dispersion, but the data also agree with
no broadening. Assuming this feature is all due to line emission
yields a line flux of $1.3^{+0.5}_{-0.6}\times 10^{-6}$ \phpspsqcm.
Since the continuum is too faint to obtain a reasonable constraint,
the equivalent width (EW) of the line is accordingly uncertain, but it
is likely to be larger than 2 keV.

The observed 2--5 keV flux is $1\times 10^{-14}$ \ergpspsqcm\
(corresponding to a 4--10 keV rest frame luminosity of $\simeq 4\times
10^{43}$\ergps), most of which originates from the line feature.
Extrapolating the 2--5 keV continuum model with $\Gamma = -0.5$
predicts source counts which are comparable to the detected counts in
the 5--10 keV image at the given exposure times. The observed 5--10
keV flux derived from the extrapolation is $2.5\times
10^{-14}$\ergpspsqcm, and the rest-frame 10--20 keV luminosity
estimated from the flux is $1\times 10^{44}$\ergps.

\section{Discussion}

\subsection{The hidden quasar nucleus}

The detection of the prominent Fe K$\alpha $ line strongly indicates
the presence of a Compton-thick AGN in IRAS F15307+3252.  The large EW
of the Fe K$\alpha$ and the deficit of X-ray emission just below the
line imply the lack of the direct continuum emission in the Fe K band,
which means the absorbing column density to be \nH $\sim 1\times
10^{24}$\psqcm\ or larger. Thus, together with the line centroid at
6.4 keV, the hard X-ray emission from IRAS F15307+3252 is dominated by
reflection.

Using the observed line luminosity, we made a rough estimate of the
minimum rest-frame 2--10 keV luminosity which is required to produce
the observed Fe K emission. Assumptions made for this calculation are
as following: Fe K$\alpha $ emission is produced in cold matter with
Solar abundance ($N_{\rm Fe}/N_{\rm H} = 3.3\times 10^{-5}$, Morrison
\& McCammon 1983), irradiated by a power-law source with a typical
quasar photon-index of $\Gamma = 2$; the continuum photons above the
K-shell ionization threshold energy (7.11 keV) are converted into Fe
K$\alpha $ at 6.4 keV with the efficiency calculated by Basko (1978)
and George \& Fabian (1991), as such that would produce $EW =150$ eV,
if an observer see both the illuminating source and reflection from
the cold slab behind it; and the solid angle of the cold matter viewed
by the illuminating source is $4\pi$.  With these assumptions, the
ratio $L_{\rm FeK}/L_{\rm 2-10}$ is found to be about 3 per cent.
Since not all the reflecting surfaces are likely visible to an
observer, the minimum 2--10 keV luminosity to produce the observed Fe
K line is estimated to be $L_{\rm 2-10} = (0.03)^{-1}\times 4\times
10^{43}\sim 1\times 10^{45}$\ergps. The ratio of bolometric to the
2--10 keV band luminosity, $f_{\rm Xbol} = L_{\rm bol}/L_{\rm 2-10}$,
is typically 30--50 for quasars (Elvis et al 1994). With this
bolometric correction, the bolometric luminosity of the AGN in IRAS
F15307+3252 is found to be $L_{\rm bol}^{\rm AGN}\geq 3\times
10^{46}f_{\rm Xbol,30}$ \ergps, where $f_{\rm Xbol}=30f_{\rm
  Xbol,30}$. Therefore a large fraction of the bolometric luminosity
($10^{47}$\ergps) is likely to be powered by the hidden AGN,
consistent with previous estimates (e.g., Yun \& Scoville 1998; Aussel
et al 1998; Verma et al 2001; Peeters, Spoon \& Tielens 2004).

The reflected X-ray emission originates from optically thick matter
and is of a distinct origin from the scattered UV/optical light
revealed by spectropolarimetry (Hines et al 1995). The large
polarization in the UV/optical bands (16--20 per cent) is presumably
due to dust scattering.

\subsection{Black holes in hidden quasars and their environment}

The black hole masss hosted in IRAS F15307+3252 is estimated to be
$M_{\rm BH}\sim 1.3\times 10^9$ \Ms, using the empirical relation
based on the virial theorem for the BLR obtained by McLure \& Jarvis
(2002) with the MgII line width of FWHM$\sim 10,000$ \kmps\ and the
3000 \AA\ luminosity ($\lambda L_{\lambda}\approx 2\times
10^{45}$\ergps) of the intrinsic UV spectrum derived from the
spectropolarimetry (Hines et al 1995). The bolometric luminosity of
this object is then about half the Eddington luminosity for the
estimated black hole mass.

IRAS F15307+3252 shares many properties with another hyperluminous
infrared galaxy IRAS 09104+4109. In their spectral energy
distribution, the infrared to optical and 60 $\mu$m to 1.5 GHz
luminosity ratios for both galaxies are around 300 and 50,
respectively (Cutri et al 1994). Both contain a hidden quasar revealed
by optical spectropolarimetry (Hines et al 1999), which appears to
dominate the energetics of the infrared emission, and their X-ray
spectra show characteristics of Compton-thick AGN (Franceschini et al
2000; Iwasawa et al 2001). The black hole mass of IRAS 09104+4109
estimated by the same technique as described above is $3\times
10^9$\Ms (V.D. Ivanov, priv comm), similar to that of IRAS
F15307+3252. These suggest that both galaxies have a well-grown black
hole emitting at a high efficiency, although their radiation is
strongly attenuated by dust shrouds in the optical wavelength and by
cold gas with a large Thomson depth ($\tau_{\rm T}\geq 1$) in the
line-of-sight in X-ray, respectively. Non-detection of CO (Evans et al
1998; Yun \& Scoville 1998) in either galaxies limits the material
needed for a vigorous star formation. The deficit of cold ($T\sim 40$
K) dust in both galaxies inferred by the non-detection with SCUBA
(Deane \& Trentham 2001; which places the limit of the contribution of
cold dust to the bolometric luminosity less than 0.3 per cent) can be
a natural consequence of the absence of a strong starburst. Perhaps in
these two galaxies, a mature black hole is still efficiently accreting
material from small radii, while star formation on a large scale has
already been terminated. As Hines et al (1995) suggested, the AGNs in
these galaxies are viewed from an unfavourable direction but otherwise
are indistinguishable from normal quasars.

One remaining similarity between the two objects is their environment.
IRAS 09104+4109 is a giant elliptical located at the centre of a rich
cluster at $z=0.44$ (Kleinmann et al 1988), which is one of the most
luminous X-ray clusters with $L_{\rm bol}^{\rm CL}\simeq 3\times
10^{45}$\ergps (Fabian \& Crawford 1995; Iwasawa et al 2001). Despite
the much lower luminosity and higher redshift, IRAS F15307+3252 appears
to be in a cluster as the extended X-ray emission indicates. If this
is the case, galaxy mergers in an overdensity region may be a
necessary condition for making a luminous quasar. This contrasts with
nearby lower luminosity infrared galaxies which are not usually found
in a rich environment (Sanders \& Mirabel 1996). On the other hand,
quasars (Wold et al 2000, 2001) and possibly hyperluminous infrared
galaxies (Farrah et al 2004) tend to reside in a moderately dense
environment, and which is in agreement with galaxy and quasar formation
model based on a hierarchical assembly scenario (e.g., Enoki,
Nagashima \& Gouda 2003). This has implications for a search for Type
II quasars in X-ray. In an unobscured quasar, the nuclear emission dominates
the total X-ray emission even if there is some emission from a cluster
with which the quasar is associated.  However, in an obscured quasar, the
nuclear emission is suppressed and soft X-ray emission from a putative
cluster could be a main X-ray source rather than the quasar itself.
Whether faint reflected light of a hidden nucleus is detectable
depends on its luminosity. Since such a component was only barely
detected in IRAS F15307+3252, one of the most luminous galaxies, with
a 30 ks XMM-Newton exposure, there are clear difficulties in detecting
Compton-thick AGN at redshift of $z\geq 1$.

\section*{Acknowledgements}

This paper is based on observations obtained with XMM-Newton, an ESA
science mission with instruments and contributions directly funded by
ESA Member States and NASA. ACF and CSC thank the Royal Society for
support. KI thanks PPARC for support.

\end{document}